# An Analysis of the New EU AI Act and A Proposed Standardization Framework for Machine Learning Fairness

Mike H.M. Teodorescu
Information School, RAISE Center, IEEE Senior Member
University of Washington
Seattle, WA, USA
miketeod@uw.edu

Yongxu Sun
Information School, MSIM Program
University of Washington
Seattle, WA, USA
yongxs@uw.edu

Haren N. Bhatia
Information School, MSIM Program
University of Washington
Seattle, WA, USA
hnb24@uw.edu

Christos Makridis
Digital Economy Lab, Stanford University, Stanford CA USA
University of Nicosia, Cyprus
makridis.c@unic.ac.cy

*Abstract*—The European Union's AI Act represents a crucial step towards regulating ethical and responsible AI systems. However, we find an absence of quantifiable fairness metrics and the ambiguity in terminology, particularly the interchangeable use of the keywords "transparency", "explainability", and "interpretability" in the new EU AI Act and no reference of transparency of ethical compliance. We argue that this ambiguity creates substantial liability risk that would deter investment. Fairness transparency is strategically important. We recommend a more tailored regulatory framework to enhance the new EU AI regulation. Furthermore, we propose a public system framework to assess the fairness and transparency of AI systems. Drawing from past work, we advocate for the standardization of industry best practices as a necessary addition to broad regulations to achieve the level of details required in industry, while preventing stifling innovation and investment in the AI sector. The proposals are exemplified with the case of ASR and speech synthesizers.

## I. Introduction

Artificial intelligence (AI) technologies have ushered us into a new era of innovation, with profound implications for society, economy, and governance. As AI systems become increasingly integrated into various aspects of daily life, the need for a regulatory framework that ensures these technologies are developed and deployed responsibly has become paramount. Already, a wide array of AI frameworks for considering the regulation of trustworthy use of AI have been introduced in the literature. The European Union (EU) has responded to this challenge with the EU AI Act (European Commission, 2021) [1], an expansive legislative effort aimed at promoting the ethical use of AI. This paper critically examines the EU AI Act, exploring its alignment with the principles of trustworthy AI and its potential impact on fairness and innovation, and proposes its blending with a set of industry standards that add more implementation detail to the EU AI Act's general frame and fairness transparency.

The EU AI Act is a comprehensive attempt to regulate AI across the EU, categorizing AI systems based on their potential risk to individuals' rights and public safety. The Act delineates four levels of risk—minimal, limited, high, and unacceptable—with corresponding regulatory requirements for each category [1]. This risk-based approach is grounded in the EU's broader strategy on AI, which emphasizes the dual objectives of fostering technological excellence and ensuring trustworthiness (European Commission, 2021) [1].

Trustworthy AI encompasses a set of ethical principles and values that guide the development and deployment of AI systems. The EU's High-Level Expert Group on Artificial Intelligence has identified seven key requirements for trustworthy AI, including transparency, fairness, and accountability [2]. These principles are echoed in studies that advocate for AI systems to be designed in ways that respect human autonomy, prevent harm, and ensure justice and equity [3]; [4].

Fairness transparency of AI products is strategically important for legal compliance, liability avoidance, and public relations. However, this issue has not been thoroughly addressed in current regulations and business strategy literature.

## II. Proposed Self-Reporting Framework

Considering the legislature over broadly defined AI is not an optimal option for detailed definition and enforcement purposes, what could be the way forward? We suggest a hybrid system with a strong focus on standardization across the industry via a centralized public platform for self-assessed yearly audits addressing fairness, human augmentation, explainability, and interpretability, while still creating a strong legislature over high-risk predictive or generative services.

Several authors have argued for some form of framework for XAI, or for transparency or interpretability of AI [48], [50]; in addition, IEEE issued a "Guide for an Architectural Framework for Explainable Artificial Intelligence'. Our contribution is to align in a single framework AI transparency and fairness, such that fairness level cand be derived in an explainable, possibly interpretable manner. In some respects, our framework has similarities with the study by Balasubramaniam et al. [52].

Partly following the literature [52], [47], we understand interpretability as the capacity of the AI to model a cause-effect relationship, where the model is readily understood by humans. Interpretability offers the clearest quantitative and qualitative clarifications to humans [47]. "Shallow" AI/ML, such as linear predictors, are fully interpretable. Explainability refers to a translation by the machine of the established relationship between inputs (data) and outputs (predictions) in a way that is well understood by humans, although the details of establishing the connection are not available. Transparency means, in this study, the entirety of methods AI conveys meanings for the results to the human user; transparency covers interpretability and explainability, but vaguer, purely qualitative representations of the causality relations established between data and predictions also pertain to transparency.

A greater challenge is, however, to make transparent and explain how and why the predictions are fair and according to what fairness metric, out of the extant large number of metrics, they obey. In turn, fairness transparency allows auditability, which is key in algorithmic fairness.

To maintain fairness, transparency, interpretability, and explainability, we emphasize displaying the quantifiable fairness metrics, i.e., fairness through awareness, demographic parity, equalized opportunity, and equalized odds.

TABLE I. LEGEND FOR AI FAIRNESS AND TRANSPARENCY METRICS DASHBOARD

| Label | Description |
|---|---|
| Disclaimer | Fairness metrics are relative; no AI is completely unbiased. |
| BOC (Bias Optimization Certificate) | Certifies trustworthiness from a standardizing body. |
| Dynamic Filters | View metrics based on protected attributes. |
| User Count | Number of users impacted by selected filters. |
| Industry Standard Metrics | Metrics for future AI applications, vendor suggested, and board approved. |
| Audit Flag | Flag to check if the company's reports have been audited by the standardizing board's auditing committee. |
| Data Snapshot Info | Includes timestamp, dataset names, and encrypted data for auditing. |

In work from two of the authors, we employed Explainable AI (XAI) techniques, such as LIME, to identify biases within the decision-making process when opaque models are utilized by AI vendors [53]. Additionally, the standardization committee can attest to the display of custom-approved metrics if required to address the nuances of future AI technologies. We believe that the standardized framework be made with industry experts and provide baseline guidance on how it can be implemented. A flag variable could be displayed to know if the company has been audited independently and would link it to the historical findings of the previous audits. Lastly, a timestamp, dataset name, and encrypted data snapshot can be maintained for external auditing. Additionally, a blockchain-based solution for fairness checks may provide more confidence to users in terms of trusting the checks and the transparency of the entire process.

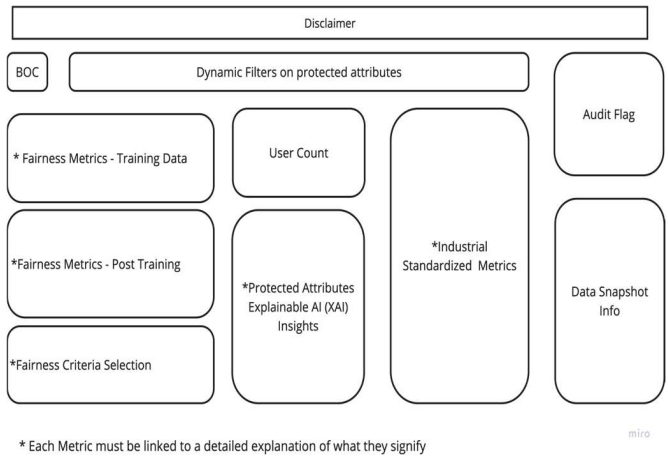

Fig. 1. AI Fairness & Transparency Dashboard Overview aimed to improve transparency of the fairness of the AI tools

Our proposed framework (see Figure 1) intends to function as a springboard for the standardization committees and the industry experts to build upon. The main elements and their motivation are briefly explained in Table 1. Fairness metrics must be applied to the data, whereas data affect what the AI system learns, and post training, for checking that the AI does not misinterpret parts of the data or distorts and bias the extracted information. The user oversees explicitly selecting the fairness criteria used, in agreement with the application. Industrial standardized metrics are more general, as they apply to all AI tools and may regard aspects less relevant to the developed tool. Auditing is mandatory to ensure that errors have not affected the development, that the AI tool is appropriate to the application and the task, and that the designers of the AI tools have not introduced involuntary biases in the data and the algorithm.

### A. High-Risk Legislature and Standardization

Understanding the limitations in defining AI and future such definitions, we believe a method of standardization would be appropriate. These standardizations would take inspiration from other industries such as the International Standard Organization (ISO) for Quality Management Standards ISO9001, the International Electrotechnical Commission (IEC), and the ISO standard for Cybersecurity ISO/IEC 27032 [5]. We call upon such organizations to These organizations should have a unified and evolving approach to best practices, base security requirements, self-auditing, and explainability approaches. Services that follow these standards will be able to advertise a badge that would represent trustworthiness to their stakeholders. We posit that these standards should approach fairness from the perspective of organizational justice theory, addressing procedural and distributive fairness. Standardization would encourage flexibility thereby promoting innovation, would be developed speedily, and facilitate international consensus to a base level [6]; [5]. However, regulations will help provide enforcement to prevent the misuse of AI and its suppression of its unaware users. Therefore, we promote a hybrid approach of base-level standardization and enforceable legislature on high-risk Artificial intelligence, like the EU AI Act but with a more restrained approach to enforcement over a narrower audience of predictive services, while still supporting innovation outside of the sandbox approach. To enforce High-Risk Artificial Intelligence, the legislature should consider the factors mentioned in Figure 1. Emboldening the importance of considering impact and having more transparency on the methodology used to pick 'High-risk' artificial intelligence would help enforcement, future proofing, and interpretability of their choices. Adding transparency to the fairness operation in the same framework would increase users' confidence and attractivity of the AI tools.

### B. Fairness Metrics

The implementation of fairness criteria within the EU AI Act is notably ambiguous. The Act mentions the importance of avoiding discriminatory impacts and unfair biases but lacks specific metrics or methodologies for achieving fairness [1]. The act does not clarify this requirement with a strong definition. This gap is critical, given the complexity of operationalizing fairness in AI and the nuances of this ever-evolving field, as demonstrated by the discussion on fairness criteria such as fairness through unawareness, demographic parity, and equalized odds. Without clear guidance on fairness implementation, there is a risk that AI systems may perpetuate existing biases. We propose the following framework for quantified fairness in machine learning and artificial intelligence. Next, we sketch a framework for quantified fairness in ML / AI.

TABLE II. FAIRNESS CRITERIA DEFINITIONS

| Fairness Metrics | Definitions |
|---|---|
| Fairness through Unawareness | This approach ignores the protected attributes in datasets. It hides the accountability of attributes for the predictor, which can be misleading [10]. |
| Fairness through Awareness | Protected attributes are not ignored but acknowledged by ensuring similar individuals are classified similarly. However, this approach does not address group-level fairness. |
| Demographic Parity | Maintains the same positive prediction rate across groups. |
| Equalized Opportunity | Ensures the same true positive rate (TP / (TP + FN)) across different groups of protected attributes. |
| Equalized Odds | Ensures both true positive rate (TP / (TP + FN)) and false negative rate (FN / (FN + TP)) across different protected attributes for higher fairness. |

The challenge in using automated decision-making systems or other forms of 'AI', as broadly defined in the EU AI act, comes down to the fairness of their models over individuals and groups, focusing on fairness across protected attributes. This paper will consider protected attributes as those defined by the U.S. Equal Opportunity Commission – race and color, religion, sex (including pregnancy, sexual orientation, or gender identity), national origin, age (40 or older), disability, and genetic information. While computational fairness is a difficult attribute to define, Morse employed organizational justice theory and addressed fairness as Distributive (fairness in outcome) and Procedural fairness (fairness in the process). To further delve into ensuring distributive and procedural fairness, there have been fairness criteria that can be quantified, namely, Fairness through Unawareness, Fairness through Awareness, Demographic Parity, Equalized opportunity, and Equalized Odds. To quantify fairness, we employ fairness criteria to provide an empirical understanding of a model's performance over protected attributes across individuals and groups.

### C. Transparency, Explainability, and Interpretability

A significant number of the high-performing algorithms are subject to the black-box effect, which means that they are inherently non-interpretable to human understanding. Article 13(1) of the EU's Artificial Intelligence Act states that "High-risk AI systems shall be designed and developed in such a way to ensure that their operation is sufficiently transparent to enable users to interpret the system's output and use it appropriately." Article (2) and (3) stressed that high-risk AI systems should be accompanied with clear instructions for use, which should also include "the characteristics, capabilities, and limitations for the performance of the high-risk AI systems". However, neither EU AI Act nor other regulations have explicitly required transparency of the fairness of AI and of the process of embedding fairness in AI tools.

TABLE III. PROPOSED DEFINITIONS OF TRANSPARENCY, EXPLAINABILITY, AND INTERPRETABILITY TABLE TYPE STYLES

| Terminology | Definition |
|---|---|
| Transparency | The accessibility and understandability of an AI system's internal mechanisms and decision processes to humans without technical background. |
| Explainability | The ability of an AI system to provide human-understandable reasons for its outputs or decisions, at least by human experts in the domain of the respective reasons or decisions. |
| Interpretability | The extent to which humans can understand the outputs of an algorithm due to the complexity of the model's design, when the humans are experts in the domain of the outputs. |

Our paper centers on transparency, particularly emphasizing its two critical components: interpretability and explainability. These two essential aspects of transparency are frequently used interchangeably, despite the notable distinctions among these terms within the field of AI [47]. Even though the terms have yet to be given a clear universal definition, the law should start to establish explicit boundaries when regulations are to be imposed accordingly. To facilitate this, we propose definitions for each term to ensure a clearer distinction (Table 2).

### D. Explainability through Explainable AI (XAI): Benefits and Risks

The law does not explicitly mention Explainable Artificial Intelligence (XAI) when addressing the interpretability of high-risk AI systems. Instead, in Article 14, the law emphasizes that high-risk AI systems should be aware of the "available interpretation tools and methods." Explainable AI (XAI) has developed as a crucial subfield to address the challenges associated with the limitations of AI models' decision-making mechanisms. XAI offers post-hoc explainability techniques for models that are not inherently interpretable to humans, such as text explanations, visual explanations, and local explanations. However, it is critical to acknowledge that many XAI tools currently available lack essential capabilities to effectively identify and address bias. This gap may contribute to a situation known as 'fairwashing,' where XAI tools incorrectly inform the researchers that a model is unbiased, while in truth, the biases persist undetected by the XAI tools. What is more, the connotation of transparency in the EU AI Act and XAI are different. The Act views transparency as a tool for promoting human rights and sustainable innovation, while XAI treats it more narrowly, aiming solely at unveiling algorithmic properties without considering its socio-technical implications, including fairness.

Another underlying problem that remains unsolved is the explainability gained from XAI approaches can hardly be quantified. The evaluations, comparisons and improvements of interpretability require quantifiable measures. Past literature overviews: on the measurement techniques developed for the goodness and accountability of the explanations offered by XAI (e.g., goodness checklist, explanation satisfaction scale, explanation trustworthiness) landed on the conclusion that more quantifiable and generalizable metrics should be developed. Therefore, if XAI is to be adopted, the law should avoid ambiguity by considering incorporating requirements for the establishment of standardized and quantifiable metrics. Similar challenges are also faced by interpretability, indicating a broad need for quantifiable metrics. To provide an example, Information Transfer Rate (ITR) measures the goodness of interpretability methods (e.g. LIME, COVAR) to intuitive human understanding. However, there are considerations to consider when developing such metrics. Firstly, different domains may necessitate distinct metrics tailored to the specific requirements of users within those domains. Secondly, while metrics are essential, they should be supplemented by qualitative human evaluations. After all, the ultimate objective is to foster understanding and trust in the model, including its fairness.

### E. 2.5 Human Oversight in High-Risk Systems

In Article 14 (1), the high-risk AI systems were required to adopt a human oversight approach to enable the individuals to "correctly interpret the high-risk AI system's output". The incentive behind this approach is salient: past studies state that human-AI augmentation is the most promising path to fairness, and human oversight can promote societal values such as human dignity, legitimacy, and accountability. People are more likely to trust an AI system if they are informed that a human is involved in the decision-making process, aligning with the AIA's objective to foster public confidence to encourage the use of AI technologies to their full potential.

Despite the incentives behind it, adding a human into the loop is not the panacea to defuse AI risks (Agarwal et al., 2024). The Act itself has pointed out the danger of over-relying on the output of high-risk AI systems (automation bias); other limitations exist. The effectiveness of the human oversight approach relies on the assumption that humans can effectively oversee algorithmic decision-making. However, humans were found to perform undesirably on their overseeing tasks). This approach would also require the system to have adequate transparency so that the human overseers have substantive material to review and audit AI tools; other factors to be considered include the type of system and the level of training the overseers received.

### III. DRAWBACKS AND EXISTING COUNTERMEASURES

If we consider the dynamic filtering criteria, we need to account for the number of scores that are reported. When considering all 7 legally protected attributes 5040 (7!) fairness metric scores need to be computed. Reporting at this granularity would be computationally intensive. Additionally, while we use random sampling of the data (Beutel et al., 2019), this approach is still costly because it excludes large portions of the data. Smaller subsets are more likely to result in a lower False Positive Rate (FPR). Moreover, determining an unbiased method of sampling remains a challenge. To address these issues, AI vendors can integrate adversarial learning and correlation regulations. The proposed solution integrates with libraries like the LinkedIn Fairness Toolkit (LiFT) to compute fairness in large-scale AI applications. LiFT is a reusable library that can be integrated with several API endpoints, by leveraging Apache Spark for general purpose cluster-computing to ensure data parallelism and fault tolerance. LiFT also provides an online service to measure model drift while monitoring fairness metrics in real-time. LiFT focuses on the following fairness metrics:

- Fairness in Training Data: KL and JS Divergences, L-p Norm and Total Variation Distances, Demographic Parity.
- Fairness in Metrics Post Model Training: Distance and Divergence metrics, Aggregate Fairness Metrics, and Differences in Model performance (across each sub-group using permutation testing).
- LiFT also provides pre-processing techniques, as well as custom metrics that may differ as per use case using User Defined Functions.

There are other ML fairness toolkits that we may consider, such as IBM's Fairness 360 tool, Microsoft AzureML Interpret, and Google's What-If, all of which work on their respective clouds or stand-alone technology and are not as flexible as LiFT.

We also recommend the reflection of explainability, and Interpretability metrics as mentioned in the previous section (e.g., Goodness checklist, explanation satisfaction scale, explanation trustworthiness), as well as the development of generalized quantifiable metrics, such as the Information Transfer Rate (ITR) to measure interpretability with extended covering of both the prediction and the fairness of it. Human-augmented feedback scoring could also be a future scope for developing new metrics in the field of explainability. All these scores should be auditable by an external body set up to track and enforce quality auditing of these self-attested scores.

### IV. CASE STUDY: ASR SYSTEMS AND SPEECH SYNTHESIZERS

The domain of speech technology, comprising automated speech recognition (ASR) and speech synthesizers, is a narrow one in AI, although there are so numerous tools commercially available. As many of these tools can be found in homes for everyday use, many in the public and industry may consider this technology mature and beyond any ethical issues. While in most cases this may be true, numerous recent studies have revealed hidden ethical issues, such as limited coverage of the dialectal speech and of languages in less developed countries, discriminatory use of names and pronunciations, unsatisfactory operation in the workplace, discriminatory operation for subjects with various medical conditions, elderly, and non-native speakers, and instances of enforced anthropological features.

This sub-field is probably too narrow to elicit a special law or even a federal regulation; instead, industry standards seem to be a good solution, but standards are lacking. Recent research has proposed multicriteria means to better characterize recognition systems in related applications (speaker recognition); similar means could be adopted for characterization of ASR systems. We suggest the development of standards for improving the ethical characterization of ASR systems and speech synthesizers based on a multi-parametric and multi-criteria evaluation that includes speakers and listeners using various dialects, from a minimal set of languages, and speakers/listeners with a large spectrum of medical conditions and ages, and under various levels of ambient noise. In addition, the standards should cover the level of diversity of using proper names and the direct and indirect use of anthropologic features.

## V. Conclusion

We overviewed the European Union AI Act as a case study, with an emphasis on fairness testing, standardization and transparency. Since the new AI Act provides broader principles but leaves many implementation details still open, we proposed a standardization approach that presses upon a framework for the self-reporting of AI services with a high level of transparency of the fairness level and visibility to protect and trust the end users by providing them a peek into the 'black box' of their AI systems. The end users must be provided with direct links to the centralized public platform where the self-reported results would be published, just before the point of usage, enhancing the EU AI Act's approach with a focus on interpretability and visibility at the point of usage. As a part of this standardized approach, all predictive services that deal with protected attributes, as previously defined, must self-audit at a pre-defined frequency, and report them on a centralized public platform.

While this will not be easily enforceable, complying companies will benefit strategically and may be given a badge similar to the CE marking, which will provide a signal of the quality of the software to users and other stakeholders. This, along with self-reported audits, would add a layer of trust between end users and enterprises.

On the legislative side, the residing commission can independently probe and audit these services and their reporting randomly. For quality self-reporting, the companies must log the metadata of the reported timestamps, and data snapshots, amongst other variables that we may consider depending upon the predictive services.